\documentstyle[prb,aps,twocolumn,epsf,floats]{revtex}

\newcommand{\beq}{\begin{equation}}
\newcommand{\eeq}{\end{equation}}
\newcommand{\bea}{\begin{eqnarray}}
\newcommand{\eea}{\end{eqnarray}}

\newcommand{\bs}{{\bf{S}}}

\newcommand{\bL}{{\bf{L}}}

\newcommand{\bx}{{\bf{x}}}
\newcommand{\by}{{\bf{y}}}

\newcommand{\etal}{{\em et al.}}

\def\tit#1#2#3#4#5{{\em #5}, {#1}{\bf #2}, #3 (#4).}

\def\prl{Phys.\ Rev.\ Lett.\ }
\def\pr{Phys.\ Rev.\ }
\def\prb{Phys.\ Rev.\ B\ }
\def\jpco{J.\ Phys.\ Cond.\ Mat.\ }

\def\jpa{J.\ Phys.\ A\ }
\def\jap{J.\ Appl.\ Phys.\ }
\def\zpb{Z.\ Phys.\ B\ }
\def\jpsj{J.\ Phys.\ Soc.\ Jpn.\ }
\def\sci{Science\ }
\def\natu{Nature\ }
\def\jsp{J.\ Stat.\ Phys.\ }

\begin{document}
\draft

\twocolumn[\hsize\textwidth\columnwidth\hsize\csname @twocolumnfalse\endcsname

\title{Magnets with strong geometric frustration} 
\author{R. Moessner}
\address{Department of Physics, Jadwin Hall, Princeton University,
Princeton, NJ 08544, USA} 
\date{Key-note theory talk of HFM-2000 Conference in Waterloo, 
Canada, June 2000}

\maketitle

\begin{abstract}
A non-technical introduction to the theory of magnets with strong
geometric frustration is given, concentrating on magnets on
corner-sharing (kagome, pyrochlore, SCGO and GGG) lattices. Their rich
behaviour is traced back to a large ground-state degeneracy in model
systems, which renders them highly unstable towards perturbations. A
systematic classification according to properties of their ground
states is discussed.  Other topics addressed in this overview article
include a general theoretical framework for thermal order by disorder;
the dynamics of how the vast regions of phase space accessible at low
temperature are explored; the origin of the featureless magnetic
susceptibility fingerprint of geometric frustration; the role of
perturbations; and spin ice. The rich field of quantum frustrated
magnets is also touched on.
\end{abstract}

]

\mbox{}
\vspace{0.cm}

\begin{figure}
\epsfxsize=3in
\centerline{\epsffile{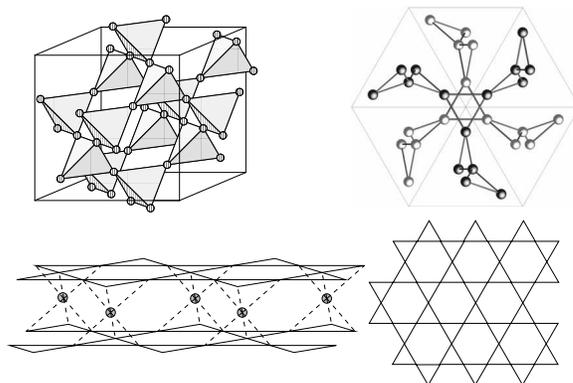}}
\caption{Corner-sharing lattices, clockwise from top left: The
pyrochlore lattice. A projection of the lattice of the Gadolinium
Gallium Garnet (GGG), which consists of two separate, interpenetrating
sublattices of corner-sharing triangles. The kagome lattice.  A
side-on view of the trilayer lattice of SCGO, consisting of triangles
and tetrahedra. It can be thought of as two kagome layers coupled by
an intermediate triangular layer (circles).}
\label{fig:latt}
\end{figure}

The concept of geometric frustration dates back to 1950, when it was
noticed that Ising antiferromagnets on the triangular lattice have
properties very different from those of ferromagnets or bipartite
antiferromagnets.\cite{wannier,houtappel} Geometric frustration has
been a topic of constant interest over the half century between then
and now. Bursts of activity have originated from developments such as
the discovery of high-temperature superconductors and the subsequent
search for unconventional magnets\cite{anderson87} or, more recently,
a still ongoing systematic study of frustrated magnetic compounds on
the highly frustrated SCGO, GGG, kagome and pyrochlore
lattices (see Fig.~\ref{fig:latt}).\cite{exptrevs,talkshfm}

Geometric frustration arises when the arrangement of spins on a
lattice precludes satisfying all interactions at the same time. The
simplest case is provided by a group of three antiferromagnetically
coupled spins: once two spins point in opposite directions, the third
one cannot be antiparallel to both of them. Geometrically frustrated
magnets are considered to be in a separate class both from
unfrustrated and from disordered magnets (spin glasses and the
like). This article concentrates on continuous, classical,
disorder-free geometrically frustrated magnetism, although discrete,
quantum and disordered models are also briefly discussed.

The popularity of geometrically frustrated magnets stems from the very
rich behaviour they present. For example, magnetic analogues of solid,
glassy, liquid and even ice phases have been identified in this class
of magnets, which is increasingly seen as providing a stage for
studying generic questions in many-body physics in a set of
well-characterised compounds described by simple model Hamiltonians. A
wide range of experimental probes are available for their study --
including neutron and X-ray scattering, muon spin rotation ($\mu$SR),
nuclear magnetic resonance (NMR), susceptibility and heat capacity
measurements -- which yield complementary information. For instance,
recently begun NMR measurements on SCGO are providing information
about the {\em local} physics at the different inequivalent sites of
the magnetic Cr ions,\cite{kerennmr98} complementing our knowledge
obtained from the probes from which such local information is harder
to extract.\cite{exptrevs}
In the following, however, only cursory reference will be made to
experiment, since a number of detailed experimental reviews exist, to
which the reader is referred.\cite{exptrevs}

\begin{figure}
\epsfxsize=3.3in
\centerline{\epsffile{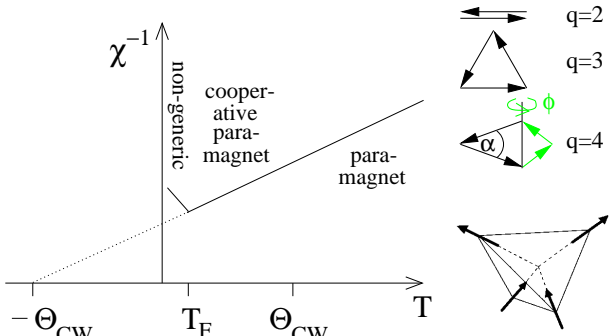}}
\caption{Left: 'Susceptibility fingerprint' of strongly frustrated
magnets.  Top right: Units of $q$\ spins with total spin
${\bf{L}}=0$. The shaded pair of spins is rotated out of the plane by
an angle $\phi$. Bottom right: The easy axes of a pyrochlore magnet.}
\label{fig:unitsofq} 
\label{fig:suscfinger}
\label{fig:spinice}
\end{figure}

Strongly frustrated magnetic compounds have a characteristic
susceptibility fingerprint (see Fig.~\ref{fig:suscfinger}). Their
inverse susceptibility, $\chi^{-1}$, follows the usual Curie-Weiss law
down to temperatures well below the expected mean-field ordering
transition temperature $\Theta_{CW}$. At some low temperature
$T_F\ll\Theta_{CW}$\ substantial deviations from the linear behaviour
occur, typically signalling a transition to a state which differs from
compound to compound, which may for example be ordered or glassy. The
smallness of the frustration parameter $T_F/\Theta_{CW}$\ has been
proposed by Ramirez to be a defining feature of `strong' geometric
frustration.\cite{exptrevs}

The regime where the temperature $T>\Theta_{CW}$\ is the usual
paramagnetic regime. The low temperature ($T<T_F$) regime is
non-generic (compound-dependent).  The intermediate regime
$T_F<T<\Theta_{CW}$, known as the cooperative paramagnetic regime,
appears to be essentially universal in this class of systems, in that
correlations remain weak although the temperature is below the scale
set by the interactions.

This observation suggests a two step strategy for understanding
magnets in this class. For the cooperative paramagnet, it should be
sufficient to study a fairly simple model system to capture the
generic behaviour characterising this regime. Building on this,
perturbations to the simple model Hamiltonian, appropriately chosen
for each compound, are introduced to describe the non-generic regime.

The remainder of this article adheres to this structure in that we
first identify and discuss an appropriate class of model cooperative
paramagnets and then consider the effect of perturbations. In the
process, we shall see that classical models of highly frustrated
magnets have in common that, once the leading, frustrated exchange
interaction has been optimised energetically, a large ground-state
degeneracy remains. The collection of degenerate ground states (the
ground-state manifold) provides no energy scale of its own and hence
any perturbation has to be considered strong. Frustrated magnets are
thus model strongly interacting systems. The richness of their
behaviour in the non-generic regime can be understood as a consequence
of the non-perturbative nature of any term added to the leading,
frustrated exchange Hamiltonian.


\section{Ground-state degeneracy of frustrated magnets}

The main distinction between frustrated and unfrustrated magnets
appears to be the presence of a large ground-state degeneracy in the
former. In the following, we first give a description of how the
degeneracy arises, and then provide a general quantitative
determination of the size of the ground-state degeneracy based on a
simple Maxwellian\cite{maxwell} counting argument.

Our starting point is the classical nearest neighbour
antiferromagnetic Hamiltonian, $H_J=\! J \sum_{\left<i,j\right>}\!
\bs_i \cdot \bs_j$, where the sum on $\left<i,j\right>$\ runs over
nearest-neighbour pairs and the spins $\bs$\ are represented by
classical vectors of unit length. $J>0$\ is the strength of the
antiferromagnetic exchange. Let us first consider the case of a group
of $q$\ mutually interacting spins, for which the Hamiltonian can be
rewritten, up to a constant, as \bea H_J=\! J \!\!\sum_{<i,j>}\!
\bs_i \cdot \bs_j\! = \frac{J}{2} \bL^2\ ,
\label{eq:hamilt} 
\eea where $\bL \equiv \sum_{i=1}^q \bs_i$\ is the total spin of the
unit.

From this it can be read off that the ground states are those states
in which the total spin $\bL$ vanishes. The appropriate configurations
for Heisenberg spins are depicted in Fig.~\ref{fig:unitsofq}. Note
that, up to global rotations, the ground states for pairs and triplets
of spins ($q=2,3$) are non-degenerate, whereas there are two degrees
of freedom, $\alpha$\ and $\phi$, in the groundstate for a quartet of
spins. The origin of this difference is the following. For Heisenberg
spins, the condition $\bL\equiv 0$\ imposes three constraints
($L_x=L_y=L_z=0$), independent of $q$. The number of degrees of
freedom, however, increases with $q$. For $q=4$, the constraints no
longer suffice to determine the ground state uniquely, and the
underconstraint shows up as the ground-state degrees of freedom.

For a lattice built up of frustrated units, this argument can be
generalised: the dimension of the ground state, $D$, is given by the
difference between the degrees of freedom, $F$, and the ground-state
constraints, $K$. Note that the Hamiltonian can be written as
$H_J=\frac{J}{2}\sum_{\alpha=1}^N \bL_\alpha^2$, where $\alpha$ runs
over all $N$\ units.

For spins with $n$ components ($n=2,\ 3$\ being XY and Heisenberg
spins, respectively), there are $n$ ground-state constraints per unit:
the $n$-component vector $\bL=0$.  The number of degrees of freedom
per unit is largest for lattices where the frustrated units share
sites, since the degrees of freedom of each spin are only shared
between two units in this case. The kagome lattice is thus made up of
corner-sharing triangles, whereas the pyrochlore lattice consists of
corner-sharing tetrahedra. In addition, more complicated lattices are
possible, for example the SCGO lattice consisting of triangles and
tetrahedra, or the GGG lattice made up of non-planar corner-sharing
triangles.

For lattices of corner-sharing units of $q$\ spins with $n$\
components, one thus obtains $D=F-K=N[n(q-2)-q]/2$.  The ground-state
dimension $D$\ grows with $n$\ and $q$. For Heisenberg spins, it
becomes extensive ($D=N$) at $q=4$, i.e. the pyrochlore
antiferromagnet has an extensive ground-state dimension. Since
physically it is hard to realise $n>3$\ or $q>4$, Heisenberg magnets
containing corner-sharing tetrahedra are the realistic systems where
the effects of frustration are strongest.

This argument has relied on the constraints being independent and
mutually compatible. For example, the ground state in a very strong
magnetic field is always non-degenerate (all spins aligned), although
the Hamiltonian can still be written as a sum of squares of
$n$-component vectors, the components of which, however, cannot all be
made zero.  Also, it turns out that the Kagome lattice Heisenberg
antiferromagnet also has $D\propto N$, because of dependent
constraints. The counting argument can thus not be applied
blindly. However, with some physical input, its simplicity can be very
useful. For the Heisenberg pyrochlore magnet, for example, $D/N=1$\
can be proven to be correct by an explicit construction of all ground
states,\cite{pyroshlo} and a careful application to the kagome magnet
can even be used to gain non-trivial insights into its behaviour under
dilution.\cite{hu} In addition, Ramirez \etal\ have explored
generalising the concept of underconstraint to a more general set of
problems, including for example the phenomenon of negative thermal
expansion as a result of underconstrained lattice degrees of freedom
(``frustrated soft mode'').\cite{ntefrus}

\section{Energy barriers and order by disorder}

\begin{figure}
\epsfxsize=3.2in
\centerline{\epsffile{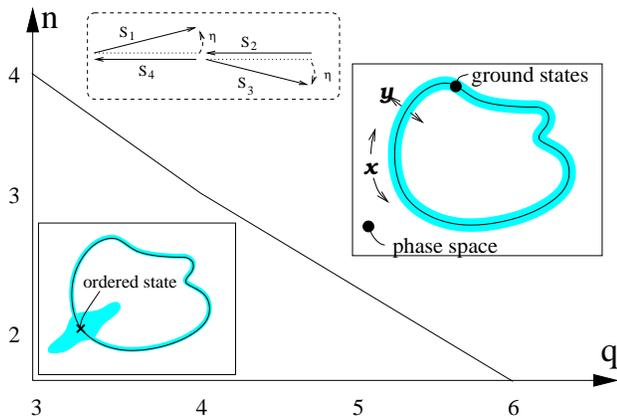}}
\caption{Schematic view of the barrier-less ground-state manifold
(black line) inside phase space. The $\bx$\ are the coordinates within
the ground-state manifold, the $\by$\ parametrise the perpendicular
directions.  The regions accessible at low temperature are
shaded. Order by disorder occurs in the region of small $q$\ and
$n$. The inset shows a 'soft mode' for a unit of four spins: this
fluctuation has zero energy cost, $E(\eta)$\ (only) in the harmonic
approximation, due to the collinearity of the spins:
$E(\eta)/J=\bL^2/2=2(1-\cos\eta)^2=\eta^4/2$.}
\label{fig:gsm}
\end{figure}

Of course, the ground state manifold has many important properties
besides its dimensionality. An important one, for example, is its
topology: can one continuously go from one ground state to any other,
or are there energy barriers separating different ground states? There
is no general answer to this question. In the case of the pyrochlore
magnets, it can be shown that there are indeed no such barriers, so
that the ground-state manifold 'comes in one piece' as depicted in
Fig.~\ref{fig:gsm}.\cite{pyroshlo}

However, at any nonzero temperature, the quantity that is minimised is
not the internal energy but rather the free energy.  Whereas the
internal energy of different ground states is the same, their free
energy can differ because of thermal fluctuations around them, which
give a different entropic weighting to each ground state. The softer
the fluctuations around a particular ground state, the larger the
region of phase space accessible near it, and the more time the system
spends fluctuating around this state. It can now happen that the
fluctuations around a special state (or set of states) are so soft
that the systems at low temperature effectively spends all its time
nearby. For this to occur, the entropy of this special set must
dominate the total entropy of all the other states taken together. In
Fig.~\ref{fig:gsm}, this is represented as the area of the shaded
region near a special point becoming much larger at low temperature
than the total shaded area elsewhere taken together.

In practice, the states with the softest fluctuations tend to
incorporate some degree of long-range order (see Fig.~\ref{fig:gsm}),
so that their selection implies an ordering transition. This
phenomenon is thus known as order by disorder, since order is induced
by thermal fluctuations, which are normally associated with a
disordering tendency (and indeed, upon raising the temperature, the
increasingly violent fluctuations do destroy the order they initially
stabilised).\cite{noorder}

The concept of order by disorder was initially proposed by
Villain\cite{villainobdo} and Shender\cite{shenderquantum}, and it has
received a great deal of attention in connection with the selection of
coplanar order in kagome Heisenberg magnets.\cite{kagch,kagplanar}
Order by disorder, although counterintuitive, has since been found to
be almost ubiquitous -- after a hard search, the first magnet shown to
avoid ordering to my knowledge was one on a Bethe
lattice.\cite{chandoucot}

In Ref.~\onlinecite{pyroshlo}, a theory based on the Maxwellian mode
counting described above was worked out to determine the presence of
order by disorder for the general case of $n$-component spins arranged
in corner-sharing units of $q$. The result is that ordered states,
provided they exist, are selected for $q$\ and $n$\ both small, as
depicted in Fig.~\ref{fig:gsm}. This region includes all the realistic
cases ($q\leq 4$, $n\leq 3$), with the exception of the case $q=4$,
$n=3$\ which is {\it marginal}. This system, the Heisenberg pyrochlore
antiferromagnet, is now universally agreed to remain disordered at all
$T$,\cite{villain,pyrodisordered,pyroshlo} a result first suggested by
Villain.\cite{villain}

This concludes the first part of our program, namely the quest for a
robust cooperative paramagnet. We can choose a classical magnet with
sufficiently large $q$\ and $n$\ and expect it to reproduce the
qualitative features of that regime faithfully, as has been done for
the Heisenberg pyrochlore antiferromagnet\cite{pyroshlo} and
subsequently for a large-$n$ kagome magnet.\cite{garanin} The
former concentrated on the low-temperature statistical mechanics and
dynamics of a cooperative paramagnet, whereas the latter contains a
detailed treatment of the thermodynamics of this regime.

\section{Dynamics}

Having established that the cooperative paramagnet can explore a vast
region in phase space even down to the lowest temperatures, the
question naturally arises how it in fact does so. This question about
its dynamics is one of the most intriguing aspects of cooperative
paramagnetism, but one which has not received its fair share of
attention over the years, despite the fact that experimental studies
of this problem are not at all uncommon.\cite{exptrevs}

The semiclassical equations of motion for a spin precessing in the
exchange field set up by its neighbours can be written as 
$ {d\bs_{\alpha\beta}}/{dt}=
 -J\,\bs_{\alpha\beta}\times({\bf L}_{\alpha}+{\bf L}_{\beta}),
$
with $\hbar=1$, and $\bs_{\alpha\beta}$\ being the spin shared by
tetrahedra $\alpha$\ and $\beta$.
This leads to a simple equation of motion for the $\bL$:
$
{d{\bf L}_{\alpha}}/{dt}=
-J\,\sum_{\beta}\bs_{\alpha\beta}\times{\bf L}_{\beta}
$.\cite{pyroshlo}

A pioneering numerical study of this dynamics was undertaken by
Keren,\cite{kerendyn} who contrasted the behaviour of Heisenberg spins
on the square and kagome lattices and who found that kagome lattice
correlations decayed qualitatively more rapidly.  This is because, in
ordered magnets, the dynamics can usually be described satisfactorily
by considering excitations around the ordered structure only, since
overall changes in the ordered structure (rotation of the ordering
direction) occur at exponentially longer timescales.\cite{andersonsym}
In cooperative paramagnets, however, the motion from one ground state
to another (parametrised by the $\bx$-coordinates), typically not
related by symmetry, can occur on relatively short timescales.

These ground-state modes have zero frequency in the harmonic
approximation (see Fig.~\ref{fig:dos}).  On top of the `spin-waves'
(locally parametrised by the $\by$-coordinates), one therefore has to
consider anharmonic effects (which are of course also of relevance for
many other properties). This was done in Ref.~\onlinecite{pyroshlo},
noticing that at low temperatures a separation of timescales
occurs. The three timescales of relevance were determined to be the
bandwidth of the excitation spectrum ($1/J$), a typical excitation
lifetime ($1/\sqrt{TJ}$), and the decay time of the autocorrelation
function ($\tau=1/T$). The dynamics looks 'diffusive': the
autocorrelation function decays like a simple exponential: $\left<
\bs(t)\bs(0)\right> =\exp(-cTt)$, where $c$\ is a constant and $t$\ is
time.


The salient feature of this result is that the decay time $\tau$\
grows only algebraically (and not, for example, according to an
Arrhenius law) as the temperature is lowered. Note that this leads to
a width, $\Gamma\propto T$, in inelastic neutron scattering linear in
$T$, which is close to what is observed in SCGO,\cite{leescgo} to
which this theory should also apply. There is no sign of
a phase transition, the cooperative paramagnetic phase extends all the
way down to zero temperature.  Note also that the excitation lifetime
is much shorter than in an ordered magnet\cite{dysonspinwave} because
the dynamics along the ground-state manifold induces a powerful
scattering mechanism.\cite{pyroshlo}

\section{The single-unit approximation and the susceptibility fingerprint}

We have seen that cooperative paramagnets have a short-range
correlations both in space and time, even in a regime where the
temperature is below the energy scale set by interactions. One can
thus hope that there might be a description in terms of variables
which behave approximately as if they were decoupled. A good candidate
set of variables are the total magnetisation vectors $\bL_\alpha$\ of
the basic units: these are the variables appearing in the Hamiltonian,
and the equations of motion (see above) can also be cast in a simple
form with their aid. In the low-temperature limit, they follow simple
equipartition, while at high temperatures, they provide a Curie-Weiss
law. These properties are shared in quantitative detail, upon
inclusion of factors of two to account for the decomposition, by the
magnet on full lattice.

\begin{figure}
\epsfxsize=3.2in
\centerline{\epsffile{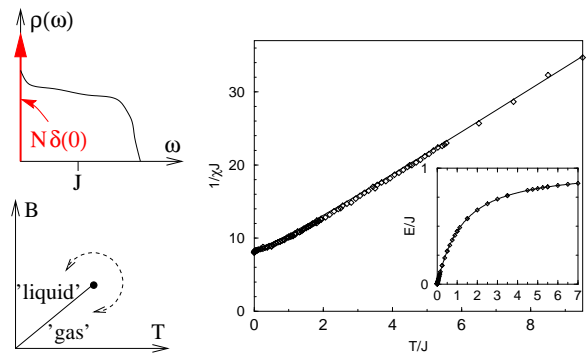}}
\caption{ Top left: The density of states of the pyrochlore
antiferromagnet in the harmonic approximation. The peak at $\omega=0$\
represents the ground-state degrees of freedom. Non-trivial dynamics
along the ground state is generated by anharmonic
interactions.\cite{pyroshlo} Bottom left: The phase diagram of spin
ice in a magnetic field along the [100] direction (after
Ref.~\cite{liquidgas}). Right: Energy (inset) and susceptibility
per spin of Heisenberg pyrochlore antiferromagnet: single-unit
approximation (line) and Monte Carlo simulations (diamonds).}
\label{fig:dos}
\label{fig:liquidgas}
\label{fig:SuscenerHFM}
\end{figure}

The partition function for isolated units of $q$\ spins were obtained
in Ref.~\onlinecite{mb}, and the pyrochlore magnet was approximated by
a set of isolated tetrahedra. In Fig.~\ref{fig:SuscenerHFM}, the
resulting expressions for energy and susceptibility are compared
against Monte Carlo simulations. The agreement is very satisfactory,
the error being below 5\% everywhere for the susceptibility and much
less than that for the energy.

Note that the shape of the susceptibility is that of the
susceptibility fingerprint pictured in Fig.~\ref{fig:suscfinger} in
that it follows the Curie-Weiss law down to temperatures well below
$\Theta_{CW}$, before bending upwards (in this case, towards the exact
$T=0$\ result for the susceptibility). The single-unit approximation
thus provides a simple model resulting in a closed-form expression for
the susceptibility of the cooperative paramagnet at all temperatures.
This model has been extended to describe quantum spins on frustrated
lattices by Garcia-Adeva and Huber.\cite{ga-h}

\section{Perturbations}

In real systems, the validity of the nearest-neighbour classical
Heisenberg exchange Hamiltonian $H_J$\ can at best be approximate. The
real Hamiltonian will symbolically have the shape $H(P)=H_J+H_P$\
where $H_P$\ denotes one of many possible `perturbation' such as
anisotropies, quenched disorder, further-neighbour exchange, dipolar
interactions, coupling to lattice degrees of freedom... If $P$\ is the
energy scale (perturbation strength) attached to $H_P$, we expect the
theory of the cooperative paramagnet to be useful in the temperature
regime $P<T<\Theta_{CW}$.

For $T\alt P$, the perturbation $H_P$\ is singular, in the following
sense. Consider, as the simplest case, a perturbation with a
non-degenerate ground state, $\Lambda$, which is at the same time a
ground state of the exchange Hamiltonian $H(0)=H_J$. This case is for
example realised, as explained in Sect.~\ref{sect:spinice}, in the
case of a pyrochlore magnet with easy axis anisotropy.\cite{anispyro}

For $P=0$, the ground-state properties are obtained from the
entropy-weighted average over the entire ground-state manifold in the
limit $T\rightarrow 0$. For an infinitesimal $P$, the ground-state
manifold (and along with it the correlation functions) discontinuously
collapses onto the state $\Lambda$.  Since $\Lambda$\ can be any one
of the multitude of states in the ground-state manifold of $H_J$,
different perturbations can result in entirely different
correlations. This in a nutshell is the origin of the richness of the
behaviour encountered in geometrically frustrated magnets.

The relationship between the ground states of $H_J$\ and of $H_P$\ can
of course be different from the simple case described above. Two other
generic scenarios are illustrated by the case of quenched
disorder. Consider first site dilution, where some ions on the lattice
are replaced by vacancies or non-magnetic ions. The corresponding
perturbation consists of shortening the spin on the site to be diluted
(until it vanishes).\cite{kagomedis} In the case of
pyrochlore\cite{villain,pyroshlo,mb,hu} and kagome\cite{kagomedis,hu}
Heisenberg antiferromagnets, at small dilution, the size of the
ground-state manifold changes but its dimension remains extensive. In
the latter case, non-coplanar spin clusters are generated by the
dilution. In the former, spins with no neighbours in one of the
tetrahedron act as uncorrelated effective (classical) spin-1/2
impurities. The presence of such a population of `orphan' spins was
first pointed out by Schiffer and Daruka in a phenomenological model
for the susceptibility of SCGO.\cite{schifferdaruka}

The other scenario is provided by bond disorder, where the strength of
the bonds has a distribution of non-zero width. Here none of the
ground states of $H_P$\ and $H_J$\ coincide. For small but nonzero
(`finite') $P$, the ground states thus do not lie on the ground-state
manifold of $H_J$, and the many ways of finding compromises between
$H_P$\ and $H_J$\ lead to a rugged energy landscape with barrier
heights of order $P$.\cite{bonddis} The existence of barriers is
necessary to account for the glassiness seen in many
compounds.\cite{gingrascrit} Recent XAFS experiments on lattice
disorder in one such pyrochlore compound
(Y$_2$Mo$_2$O$_7$)\cite{booth00} do suggest a distribution of bond
lengths wide enough to give rise to a substantial degree of bond
disorder.

\section{Quantum frustration}

Of all perturbations, the introduction of quantum fluctuations is
probably the most interesting as the ground-state wavefunction can
turn out to be any linear combination of the classical ground
states. Unusual correlated or disordered (`spin liquid') magnetic
states, with unconventional excitations and quantum phase transitions,
can thus arise.

A gentle approach to quantum magnetism lies in a semi-classical
(large-$S$) treatment of our model Hamiltonian $H_J$. The leading
effect in $1/S$\ is the generation of a zero-point contribution to the
effective energy of a classical ground state,\cite{shenderquantum}
which looks like a classical perturbation. This energy may be
represented by a bilinear term favouring collinearity in an effective
energy functional.\cite{henleyvec} However, the harmonic analysis at
$O(1/S)$\ may still preserve a massive degeneracy, as happens in the
case of the kagome magnet, where all the coplanar states have the same
zero-point energy.\cite{kagch} Selection of a single state in that
case is believed to be caused by anharmonic interactions, which select
a $\sqrt{3}\times\sqrt{3}$\ configuration with a tripled unit
cell.\cite{kagan,sachtri} Similarly, a biquadratic term favouring
collinear configurations retains the extensive zero-point entropy of a
pyrochlore Ising antiferromagnet.\cite{Liebmann}

As the spin length is decreased further towards $S=1/2$, the strength
of the quantum fluctuations increases. Sufficiently violent quantum
fluctuations might destabilise any ordered structure present at large
$S$, the same way that strong thermal fluctuations destroy thermal
order by disorder. There is strong evidence from numerics that the
$S=1/2$\ kagome Heisenberg antiferromagnet is indeed quantum
disordered,\cite{kagdiag} as are the kagome Ising antiferromagnet with
quantum fluctuations introduced via a transverse
field\cite{kagtrfield} or the triangular lattice Heisenberg magnet
either with a multiple-spin exchange term added to increase the
strength of quantum fluctuations,\cite{trimult} 
in a ``large-N'' treatment,\cite{sachtri}
or in a valence bond
dominated phase.\cite{trirvb} Similarly, a perturbative
analysis by Canals and Lacroix of the $S=1/2$\ pyrochlore
antiferromagnet finds a quantum-disordered phase.\cite{canalspyro} For
this case, however, Harris \etal\ have suggested that
long-range order may nonetheless be present in higher-order spin
correlation functions,\cite{harber} and Isoda and Mori have proposed
the existence of a valence-bond crystal.\cite{isodapyro}

As indicated above, there is much more to quantum frustration than the
discovery of such quantum spin liquids.  However, a proper discussion
of quantum frustration lies beyond the scope of this article, and we
now move on to quite a different subject, the magnetic version of ice.

\section{Spin Ice}
\label{sect:spinice}

Probably the most remarkable recent experimental development has been
the discovery of `spin ice', in experiments on the compound
Ho$_2$Ti$_2$O$_7$,\cite{harspinice} which was supplemented by another
titanate compound, Dy$_2$Ti$_2$O$_7$.\cite{ramspinice}

Harris \etal\ noticed that a strongly anisotropic {\em ferromagnet} on
the pyrochlore lattice is frustrated, whereas an antiferromagnet is
not. The reason for this lies in the orientation of the easy axes,
depicted in Fig.~\ref{fig:spinice}, which the spins are constrained to
point along by the anisotropy. The exchange Hamiltonian
(Eq.~\ref{eq:hamilt}) seeks to mimimise (maximise) the total spin of
the tetrahedron in case of antiferromagnetic (ferromagnetic)
exchange. The two configurations with the spins pointing all in and
all out on alternating tetrahedra have $\bL_\alpha=0$\ everywhere.
The antiferromagnetic ground state is thus only trivially degenerate
and appears unfrustrated. By contrast, there are exponentially many
configurations maximising the total spin on each tetrahedron, namely
all those with two spins pointing in and two pointing out on each
tetrahedron. This is equivalent to the ice model, as the ice rules
state that each oxygen has two protons sitting near and two far from
it in the ice structure. In this model, a spin pointing out/in is
taken to represent a proton sitting near/far from an imaginary oxygen
placed at the centre of the tetrahedron.\cite{harspinice} These ice
states are in fact the ground states of an Ising model on the
pyrochlore lattice, which curiously had already been noticed by
Anderson\cite{andersonpyro} in the first discussion of frustration on
the pyrochlore lattice in 1956.  Thus we now have a magnetic compound
which cannot only be used to study ice physics but which also turns
out to have a large number of other interesting
aspects.\cite{harspinice,ramspinice,andersonpyro,icerm,icebh,liquidgas,gregice,icesid,icegh,icebymi}

Theoretical work has so far thrust in two directions. 
Firstly, unusual properties of spin ice as a model many-body system
are being explored.  For example, the spin ice ground states are
massively (discretely) degenerate, giving the system an extensive
zero-point entropy.\cite{Liebmann} Harris \etal\cite{liquidgas} noted
that a magnetic field, $B$, applied in the [100] direction can lift
this degeneracy completely as a result of the orientations of the easy
axes. The contributions of both the field and the entropy to the free
energy are extensive, the latter being weighted by the temperature. At
$T=0$, the magnetic field energy thus dominates, and the ground state
(termed an entropy-poor, ``liquid'' state) has the maximal
magnetisation compatible with the easy axis constraints.  However, as
$T$\ is increased, the entropic contribution eventually dominates, and
a first order transition to the entropy-rich (``gas'') state ensues,
with a discontinuous drop in the magnetisation.

The curious feature of this transition is that it takes place between
two states not differing in symmetry. In particular, the line of first
order transitions in the $B-T$\ plane terminates in a critical point,
very much the same way as happens in a conventional liquid-gas phase
diagram, allowing a continuous path from the liquid to the gas without
encountering any transition (Fig.~\ref{fig:liquidgas}).

Secondly, two groups are studying the microscopic details and the
resulting behaviour of the two titanate compounds mentioned
above. Both are incorporating long-range dipolar interactions in
addition to the nearest neighbour
exchange.\cite{icesid,icebymi,dipoles} Siddharthan \etal\ find
Ho$_2$Ti$_2$O$_7$\ to be in a partially ordered state,\cite{icesid}
whereas den Hertog \etal\ conclude it to be a bona fide spin ice
compound.\cite{icebymi} The origins of this disagreement are not
entirely clear and may be due to problems involved in approximating
the long-range nature of the dipolar interactions.

\section{What's not here}
In this article, I have tried to give a non-technical introduction to
and a short overview of the theory of strongly frustrated magnets. I
hope to have conveyed to the reader the idea that this field is a rich
one, and, if nothing else, that a review article not constrained by
size limits is by now overdue. I had to skip many exciting topics;
these include the unconventional heavy fermion behaviour in
LiV$_2$O$_4$,\cite{lithvan} interactions of orbital and spin degrees
of freedom,\cite{znv,mgv,zncr} rigorous results on ground states of
frustrated quantum magnets\cite{liebschupp} and the many facets of
quantum itinerant magnetism\cite{ymn,lithvan} to name just a few. I
have completely omitted any mention of one-dimensional systems.  Other
overview articles can be found in Refs.~\onlinecite{diep,shender}. For
magnets on the triangular lattice, there exists a very thorough review
by Collins and Petrenko.\cite{collinsrev} Classical frustrated Ising
magnets on a wide range of lattices are reviewed in
Ref.~\onlinecite{Liebmann}.

\section*{Acknowledgements}
I am immensely grateful to John Chalker, who introduced me to this
field. Many of the ideas presented here have originated with
him. Other theorists who have generously shared their knowledge with
me over the years are John Berlinsky, Premi Chandra, Michel Gingras,
Chris Henley, Peter Holdsworth, David Huse, David Sherrington and
Shivaji Sondhi. I am also very grateful for the large number of
discussions with members, past and present, of the experimental groups
radiating outwards from Bell, Edinburgh, Grenoble, McMaster, RAL and
Vancouver. Thanks are due for figures to Mark Harris (single
tetrahedron), Oleg Petrenko (GGG) and Martin Zinkin (pyrochlore
lattice).


\begin{references}
\bibitem{talkshfm} Several of these experiments, as well as related
theoretical work, are treated in other articles of this
volume.
\bibitem{noorder} It is assumed here that the projection onto the
ground state by itself does not induce any long-range order.
\bibitem{dipoles} Dipolar interactions are also being studied in the
context of experiments on Gadolinium Titanate.\cite{gdtiraju,gdst}
\end{references}
\end{document}